\documentclass[final, grl]{agutex}
\pdfoutput=1
\usepackage[dvips]{graphicx}
\usepackage{epsfig}
\usepackage{lineno}
\usepackage{color}
\usepackage{epstopdf}
\usepackage{amsmath, amssymb}
\graphicspath{ {img/} }
\usepackage{bm}
\usepackage{siunitx}
\newcommand{\planss}  {{Planatary Space Science }}  
\newcommand{\ssr}{    {Space Sci. Rev. }}

\authorrunninghead{XU ET AL.}
\titlerunninghead{Field-aligned currents at $\sim 60\,R_E$}
\authoraddr{Sixue Xu, Department of Geophysics and Planetary Science, University of
 Science and Technology of China, Hefei, China (hzxusx@mail.ustc.edu.cn)}
\authoraddr{Andrei Runov, Institute of Geophysics and Planetary Physics,
University of California, Los Angeles, USA}
\authoraddr{Anton Artemyev, Institute of Geophysics and Planetary Physics,
University of California, Los Angeles, USA}
\authoraddr{Vassilis Angelopoulos, Institute of Geophysics and Planetary
Physics, University of California, Los Angeles, USA}
\authoraddr{Quanming Lu, CAS Key Laboratory of Geospace Environment, Department of
 Geophysics and Planetary Science, University of Science and Technology of China, 
Hefei, China (qmlu@ustc.edu.cn)}

\begin{document}
\title{Intense cross-tail field-aligned currents in the plasma sheet at lunar distances}
\author{Sixue Xu \altaffilmark{1,2}, Andrei Runov \altaffilmark{1}, Anton
Artemyev \altaffilmark{1,3}, Vassilis Angelopoulos \altaffilmark{1},  and
Quanming Lu \altaffilmark{2}}
\altaffiltext{1}{Institute of Geophysics and Planetary Physics, University of
California, Los Angeles, USA}
\altaffiltext{2}{CAS Key Laboratory of Geospace Environment, Department of
Geophysics and
Planetary Sciences, University of Science and Technology of China, Hefei, China}
\altaffiltext{3}{Space Research Institute, RAS, Moscow, Russia}

\begin{abstract}
Field-aligned currents in the Earth's magnetotail are traditionally associated
with transient plasma flows and strong plasma pressure gradients in the
near-Earth side. In this paper we demonstrate a new field-aligned current system present at the lunar orbit tail. Using
magnetotail current sheet observations by two ARTEMIS probes at $\sim60 R_E$,
we analyze  statistically the current sheet structure and current density distribution closest to the neutral sheet.
For about half of our 130 current sheet crossings, the equatorial magnetic field component across-the tail (along the main, cross-tail current) contributes significantly to the vertical pressure balance. This magnetic field component peaks at the equator, near the cross-tail current maximum. For those cases, a significant part of the tail current, having an intensity in the range 1-10nA/m$^2$, flows along the magnetic field lines (it is both field-aligned and cross-tail). We suggest that this current system develops in order to compensate the thermal pressure by particles that on its own is insufficient to fend off the lobe magnetic pressure.
\end{abstract}
\begin{article}

{\it Main point \#1: Statistics of the magnetotail current sheet properties at
the lunar orbit }\\

{\it Main point \#2: Significant contribution of the magnetic field shear to the pressure balance for $\sim 50$\% observed current sheets }\\

{\it Main point \#3: Intense field-aligned currents ($1-10$ nA/m$^2$) at the
lunar orbit magnetotails.}\\

\section{Introduction}
Dynamics large-scale plasma systems, planetary magnetotails, is significantly determined by configuration of the magnetotail current sheets, regions with strong plasma currents \citep[e.g.,][and references therein]{Eastwood15:ssr}. The Earth magnetotail current sheet, most accessible for in-situ spacecraft investigations, represents the high-$\beta$ region where the plasma pressure significantly exceeds the magnetic field pressure. Models describing the such configuration predicts strong cross-tail currents (from dawn to dusk) flowing across the magnetic field \citep[see, e.g.,][and references therein]{Birn04, Zelenyi11PPR, Sitnov&Merkin16}. This concept is based on an assumption of the diamagnetic nature of magnetotail currents. Strong deviations from this nominal geometry and formations of cross-tail currents flowing locally along magnetic field lines were observed in the near-Earth plasma sheet during flapping events \citep[e.g.,][]{Petrukovich08} and during current sheet thinning \citep{Artemyev17:grl:currents}. Yet, the geometry of currents in the mid-distant
magnetotail remains unknown.

At the lunar orbit the Earth dipole magnetic field is vanishing and the
average positive north-south component $B_z$ (GSM coordinates are used through
the paper) is smaller than $B_z$ fluctuations. The lack of the
dipole-dominated magnetic field configuration \citep[the lunar orbit magnetotail
is more influenced by solar wind conditions than by Earth dipole filed, see,
e.g.,][and references therein]{Sibeck&Lin14} and the presence of transient
intense currents \citep[e.g.,][]{Pulkkinen93, Vasko15:jgr:cs} make the
magnetotail current sheet very dynamical and unstable. Therefore, one can expect
an existing of more complicated and exotic current sheet configuration in the
mid-distant tail in comparison with the near-Earth controlled by
the Earth dipole field. An important question is how global magnetotail dynamics
(e.g., rotation of the equatorial plane following the interplanetary magnetic
field, large-scale flapping waves, etc.) deform the current sheet and can these
deformations result in new current sheet configuration including field-aligned currents?

To address this question, we investigate the current sheet properties at the
lunar orbit ($\sim-60 R_E$) using two ARTEMIS (Acceleration, Reconnection,
Turbulence, and Electrodynamics of the Moon's Interaction with the Sun) probes.
Being interested in local current sheet structure, we focus on flapping current
sheets and use plasma velocity to reconstruct the spatial scales and
current densities \citep{Sergeev98}. Collected statistics of the current sheet
observations reveal previously unknown property of this magnetotail region: about half of observed sheets are characterised by significant contribution of the magnetic field shear to the pressure balance. Such atypical for the Earth magnetotail current sheets have a local equatorial peak of the magnetic field component directed along the current density and the corresponding intense cross-tail field-aligned currents with magnitudes $\sim 1-10$ nA/m$^2$. Similar current sheet configurations are frequently observed in solar wind
\citep[e.g.,][]{Paschmann13:angeo} and planetary magnetotails where plasma
pressure can be insufficiently high to support the pressure balance \citep[e.g.,
Jupiter, Venus, Mars, see][]{Artemyev14:pss,Rong15:venus,Artemyev17:jgr:mars},
but in the near-Earth magnetotail such current sheet configuration were found
only within close vicinity of the reconnection region
\citep[e.g.,][]{Nakamura08}. In contrast, we have demonstrate that at the lunar
orbit the current sheets with intense cross-field field-aligned currents
are very representative state of the magnetotail current sheet.

\section{Data Set and Analysis Technique}
Since July 2011 both ARTEMIS probes (P1 and P2) has been in stable equatorial,
high-eccentricity, 26-hr period orbits $\sim$100\,km x 19,000\,km altitude over
the Moon. The inter-probe separations varies between 500\,km and 5\,R$_E$. The
ARTEMIS duo traverses the magnetotail during about four days per a month. In
this study we use measurements of the ARTEMIS fluxgate magnetometer \citep[16
vectors per second in fast survey, see][]{Auster08:THEMIS} and ion moments
provided by combined data of the electrostatic analyzer \citep[ESA, energies
below $\sim 25$ keV energies][]{McFadden08:THEMIS} and solar state telescope
\citep{Angelopoulos08:sst}.  For electron moment we use only ESA measurements.
We use only fast survey data, what determines the dominance of observations in
the duskside ($Y_{GSM} >0$).

For one year of ARTEMIS observation (from June 2016 to June 2017) we visually
select all current sheet crossings (130 events) characterized by $B_x$ reversal
within less than 20 min and with the $B_x$ span larger than 5 nT. To determine
the local coordinate system, we use the minimum variance analysis (MVA)
resulting in the $\bm{l}$, $\bm{m}$, and $\bm{n}$ vectors of maximum, medium,
and minimum variances, respectively \citep{Sonnerup68}. The maximum variance
direction is mainly along the sign-changes component ($B_x$),  the minimum
variance direction is treated as the normal to a plane current sheet (i.e.,
direction of the main gradient), and the intermediate variance direction is
along the nominal electric current direction.

\begin{figure}[h]
\centering
\begin{tabular}{c}
\includegraphics[width=0.43\textwidth, trim={3cm, 0cm, 7cm, 0cm},
clip]{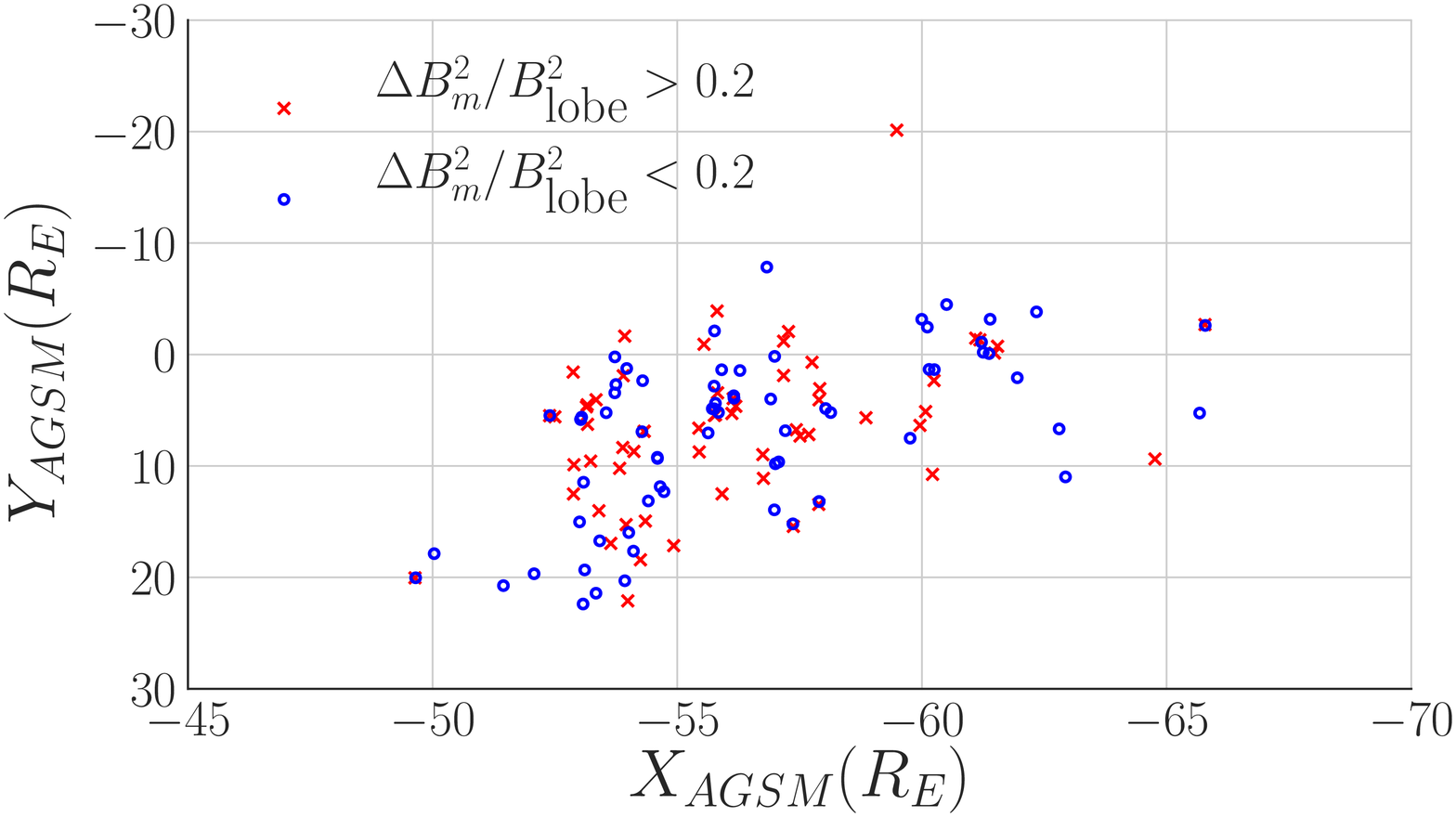} \\
\includegraphics[width=0.43\textwidth, trim={5cm, 0cm, 4cm, 0cm},
clip]{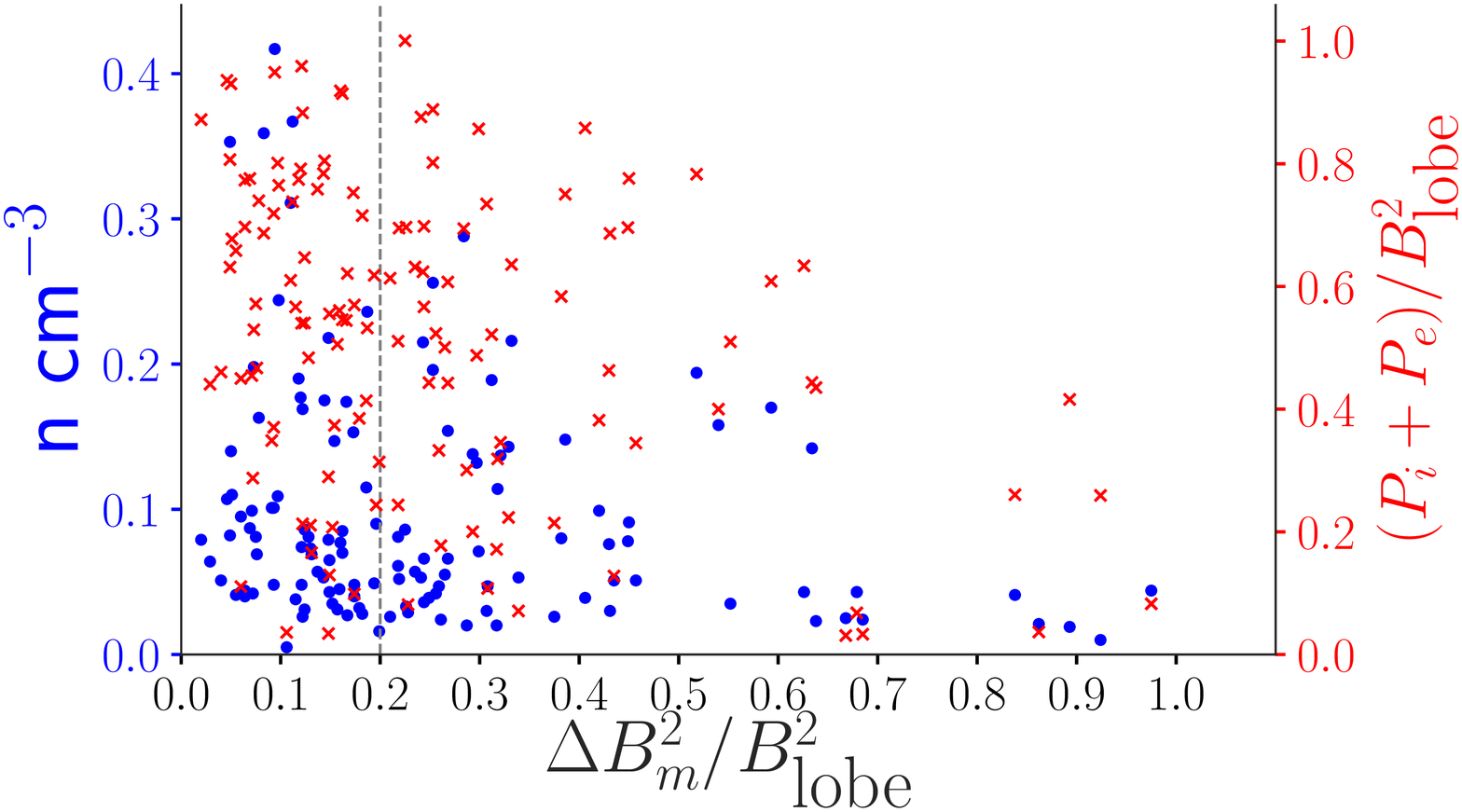}\\
\includegraphics[width=0.43\textwidth, trim={3cm, 0cm, 7cm, 0cm},
clip]{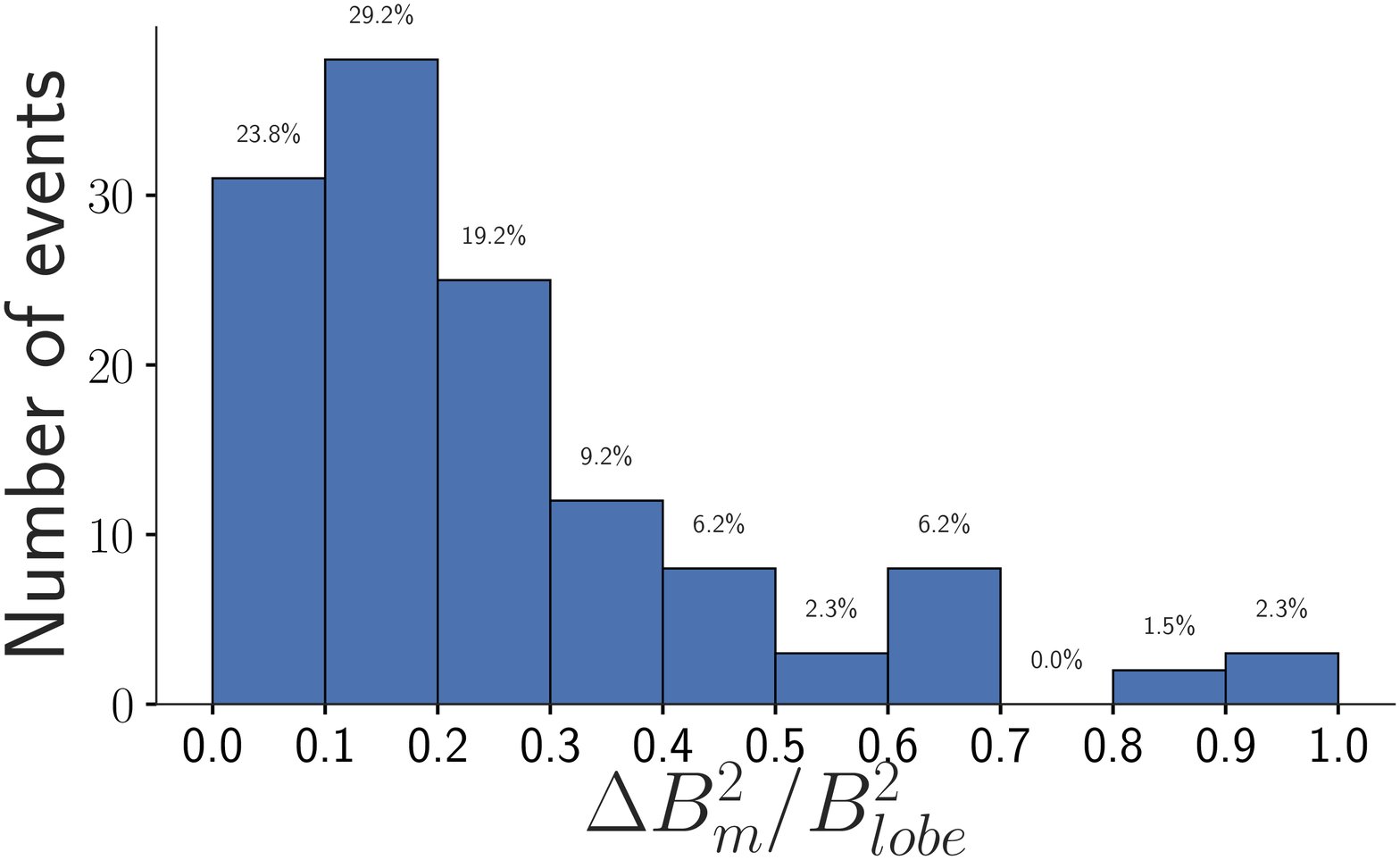} \\
\end{tabular}
\caption{(a) distribution of current sheets with $\Delta B_m^2/B_{\rm
lobe}^2<0.2$ (blue circles) and $\Delta B_m^2/B_{\rm lobe}^2>0.2$ (red crosses).
(b) plasma density and thermal pressure in the current sheet center
$|B_l|<0.4B_{\rm lobe}$ nT as functions of $\Delta B_m^2/B_{\rm lobe}$. (c)
histogram of $\Delta B_m^2/B_{\rm lobe}^2$ distribution. }
\label{fig:distribution}
\end{figure}

We have checked that for all selected currents sheets the pressure balance
$B_l^2+B_m^2+2k_B\mu_0n_e(T_i+T_e)=B_{\rm lobe}^2\approx const$ is satisfied
within $20$\% (here $n_e$, $T_i$, $T_e$ are electron density, ion and electron
temperatures).  The pressure balance analysis reveals two distinct categories of
current sheet structure: 1) currents sheets where the magnetic pressure $B_{\rm lobe}^2\gg
B_m^2$ is fully balanced by the thermal plasma pressure $2k_B\mu_0n_e(T_e+T_i)$,
and 2) current sheets with large contribution of $B_m$ variation to the pressure
balance. In the latter cases, $B_m^2$ peaks around the equatorial
plane $B_l=0$ and the $B_m^2$ variation across the sheet $\Delta B_m^2$
comparable to $\Delta B_l^2\approx B_{\rm lobe}^2$. Figure
\ref{fig:distribution}(a) shows spatial (in $(X,Y)$ GSM aberrated on $4^\circ$)
distribution of current sheets with $\Delta B_m^2/B_{\rm lobe}^2<0.2$ and
$\Delta B_m^2/B_{\rm lobe}^2>0.2$ (with $B_m \gtrapprox 0.4 B_{\textrm{lobe}}$).
A significant
population of selected current sheets are characterized by $B_m$ contribution to
the pressure balance. This is distinctive feature of the middle (lunar orbit)
magnetotail, whereas in the near-Earth tail the current sheets with $B_m
\gtrapprox 0.4  B_{\textrm{lobe}}$ are observed quite rare \citep[see few
examples in][]{Nakamura08, Rong12:by}.

Figure \ref{fig:distribution}(b) illustrated the correlation between electron
number density (blue point), normalized thermal pressure
($2k_B\mu_0n_e(p_i+p_e)/B^2_{\rm lobe}$ , red cross) and $\Delta
B_m^2/B_{\textrm{lobe}}^2$. Most of events are concentrated in the lower
diagonal side, i.e. current sheets with large $\Delta B_m^2/B_{\textrm{lobe}}^2$
are characterized by low density (low plasma pressure), whereas current sheets
with large density (thermal pressure) are characterised by small $\Delta
B_m^2/B_{\textrm{lobe}}^2$. As showed in Fig. \ref{fig:distribution}(c), among
the 130 events, half of them fall in to the first category with insignificant
$B_m$ contribution to the pressure balance, and the other half have large
$\Delta B_m$ showing a long-tail distribution of $\Delta
B_m^2/B_\textrm{lobe}^2$ ranging from $0.2$ to $1.0$.

\section{Current sheet internal structure}

To investigate the inner structure of the current sheets with
large $\Delta B_m^2/B_\textrm{lobe}^2$, we select five cases with the $v_z$ (or
$\bm{v}\cdot\bm{n}$) velocity well correlating with $dB_l/dt$ that enable us to
use the reconstruction technique \citep{Sergeev98}. Table \ref{tab:cases} shows
main parameters of these current sheets (events A-E). For all events, except A,
the ratio of eigenvalues of $\bm{m}$ and $\bm{n}$ vectors are large enough to
distinguish well between $\bm{m}$ and $\bm{n}$ directions. Interplanetary
magnetic field $B_y$ (IMF) measured within 30 minutes including the time
interval of the current sheet observations is relatively small ($\leq 1$ nT).

\begin{table*}
	\resizebox{\textwidth}{!} {%
		\begin{tabular}{ccc ccc ccc ccc cc}
			\textrm{Case}                    & \textrm{date}
           & \textrm{time}
			& $\lambda_m/\lambda_n$            & $\Delta
B_m^2/B_\textrm{lobe}^2$ & IMF $B_y$
			& $B_{\textrm{lobe}}$              & $N_i$
			& $p_e+p_i$                        & $j_l$
             & $j_m$
			& $L$                              & $L/\rho_i$\\ \hline
			&                                  &
			&                                  &
             & \si{nT}
			& \si{nT}                          & \si{cm ^{-3}}
			& \si{nT^2}                        & \si{nA/m^2}
             & \si{nA/m^2}
			& \si{km}                          &
			\\
			\hline
			
			A &2017-04-13 & 01:02:38 & 2.149 & 0.253 & -1.27 & 7.30
& 0.20 & 47.36 & 0.45 & 1.25 & 4895 & 11.05 \\
			B &2017-01-11 & 18:37:00 & 15.000 & 0.552 & -0.97 & 8.78
& 0.04 & 39.28 & -9.06 & -12.16 & 492 & 0.76 \\
			C &2017-01-11 & 11:40:33 & 7.042 & 0.593 & -0.45 & 8.76
& 0.17 & 46.70 & 2.50 & -2.05 & 1721 & 3.90 \\
			D &2016-11-13 & 19:48:18 & 16.601 & 0.317 & 0.76 & 11.85
& 0.02 & 24.06 & 8.06 & -13.53 & 516 & 1.17 \\
			E &2016-10-15 & 08:08:06 & 6.322 & 0.634 & -6.68 & 14.46
& 0.14 & 92.75 & -11.09 & 8.90 & 689 & 1.91 \\
			F &2017-01-11 & 10:45:16 & 1.864 & 0.02 & 0.60 & 9.66 &
0.08 & 81.37 & 0.10 & -0.69 & 5404 & 8.07 \\
			
			\hline
		\end{tabular}
	}
	\caption{characteristic parameters of selected crossings}
    \label{tab:cases}
\end{table*}

Using the linear
regression between $dB_l/dt$ and $v_z$, we estimate the current densities $j_m
\sim (dB_l/dt)/v_z$, $j_l \sim (dB_m/dt)/v_z$ and current sheet spatial scale
(thickness, $L$). Table \ref{tab:cases} shows that $j_m$ can
reaches $10$\,nA/m$^2$ \citep[quite large value even in comparison with the
near-Erath current sheet, see review][]{Petrukovich15:ssr}. Current sheet
thicknesses are about $500-1000$\,km and comparable with the ion gyroradius
calculated in $B_{\rm lobe}$. Thus, we deal with thin current sheets. The
magnitude of $j_l$ current is comparable with $j_m$ magnitude, what makes
selected current sheets different from the typical magnetotail current sheets
with $j_l\sim 0$. For a comparison, we also select one
example of flapping current sheet with well correlated
$v_n$ and $dB_l/dt$, but without significant $\Delta B_m^2$ (event F from Table
\ref{tab:cases}).

Figure \ref{fig:B-t} shows three sets of six panels with current sheet
characteristics. The top panels demonstrate time series of magnetic field
components (gray curves show well established pressure balance, $B_{\rm
lobe}\approx const$). There is clear $B_m$ peak around the equatorial plane
$B_l=0$ for events A-E. In all cases, $n_y$ component is significant, i.e. we
deal with tilted current sheets \citep{Zhang02}, where current density flow
partially along the north-south direction, whereas $B_m$ component is
contributed both by GSM $B_y$ and $B_z$. Although, the near-Earth observations revealed strong
field-aligned currents in the tilted current sheets \citep[e.g.,][and references
therein]{Petrukovich15:ssr}, these currents were generated due to $j_m\sim j_z$
projection to uniform $B_m\sim B_z$ field, whereas strong peak of $B_m\sim B_z, B_y$
has not been reported yet.

Middle panels of Fig. \ref{fig:B-t} show profiles of plasma and $\Delta B_m^2$
pressures versus $B_l$. For five selected current sheets the $B_m$ contribution
to the pressure balance is comparable with the thermal plasma contribution.
Variation of the plasma pressure across the current sheets is supported both
by density and temperature variations (see bottom panels of  Fig. \ref{fig:B-t}).  Density distributions of
Case A, C, E show a decease of $\sim50\%$, $\sim 30 \%$, $\sim 25 \%$ from the
maximum value at $B_l=0$. In Case A, C, F, the temperature has almost uniform
profile across the sheet, whereas in Case B, D, E, the temperature decreases to
50\% of its maximum value at $B_l=0$. The temperature
increase in case C at the current sheet boundary is due to very rarefied hot
field-aligned ion flows. It is interesting to note strong density variations
across the sheet in cases A, E and F. In the near-Earth magnetotail,
temperature variations across the sheet, $T_i(B_l)$, are usually stronger
than the density variation \citep[e.g.,][]{Runov06, Petrukovich15:ssr}.

\begin{figure*}
\centering
\includegraphics[width=0.7\textwidth, trim={6cm, 3cm, 6cm, 3cm},
]{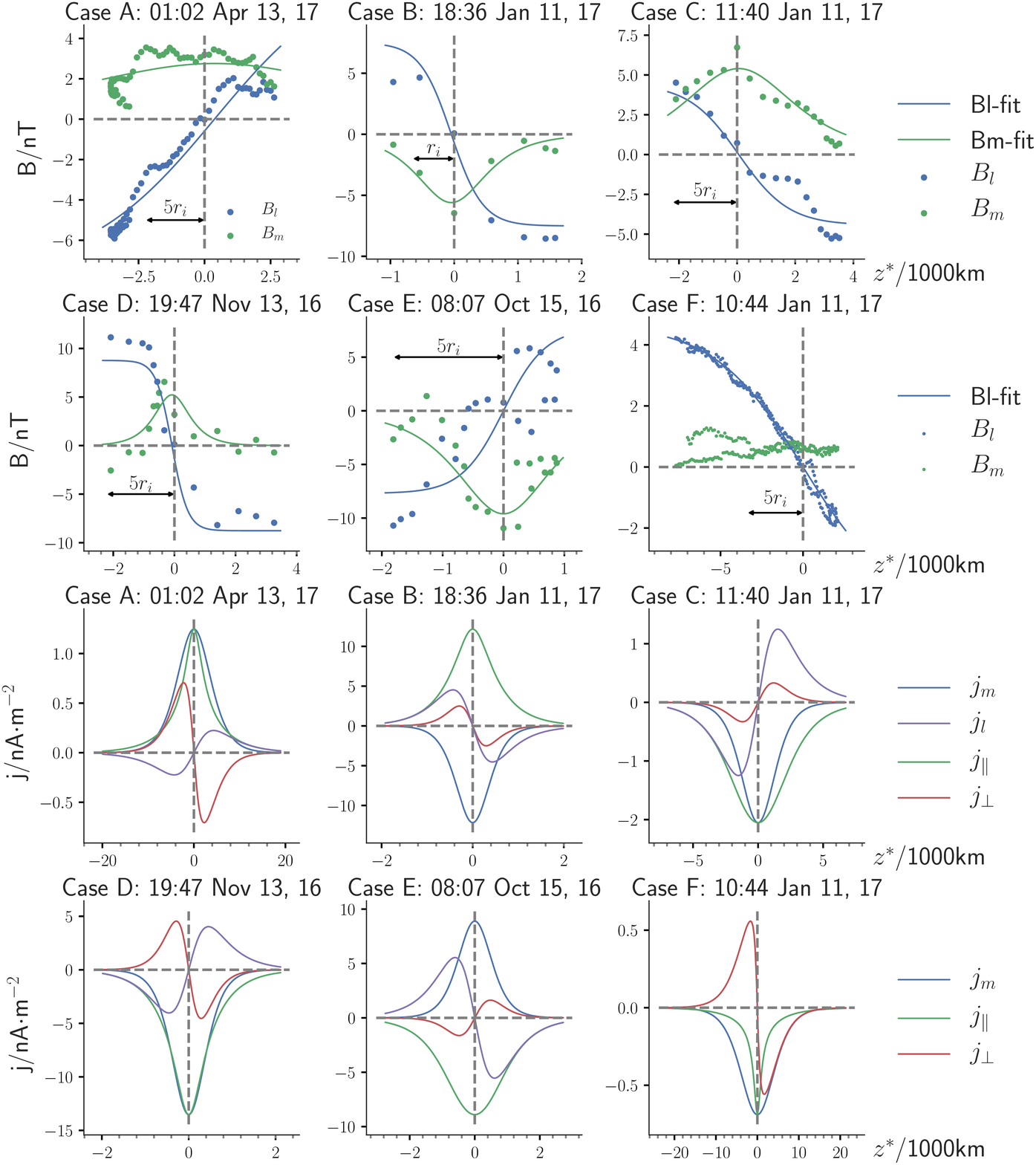}  \\
\caption{Top six panel shows the reconstructed spatial profile of $B_l, B_m$,
whereas bottom six panels show $j_l, j_m, j_\parallel, j_\perp$ profiles.}
\label{fig:B-j-z}
\end{figure*}

To reconstruct the internal current sheet structure, we adopt methods used by
\citet{Sergeev98}, \citet{Vasko15:jgr:cs} and determine time interval with good
correlation between derivative $dB_l/dt$ of the 16s smoothed magnetic field and
$v_n$(or $v_z$) velocity. Then we integrate velocity (excluding offset $v_0$
defined from $dB_l/dt=\mu_0j_mv_z+v_0$ fitting) along this interval to determine
$z$ coordinate ($z=0$ is defined at $B_l=0$). We fit $B_l(z)$ and $B_m(z)$ dependencies by
simple function $\tanh(z/L)$, $\cosh^{-1}(z/L)$ \citep[see model
in][]{Harrison09:prl}. Top six panels of Fig. \ref{fig:B-j-z} shows these
magnetic field profiles. For events A-E the $B_m$ component demonstrates the
clear bell-shaped profile, $B_m \sim \cosh^{-1}(z/L)$, with the current sheet
thickness $L$ about few ion gyroradii.

Using fitted magnetic fields, we estimate the current density profiles. Bottom
six panels in Fig. \ref{fig:B-j-z} shows strong peak of $j_m \sim 1-10$ nA/m$^2$
and bi-polar $j_l$ profiles for events A-E. The event F is characterized by
rather weak $j_m\sim 0.5$ nA/m$^2$ \citep[this is typical current density
magnitude at such distances, see][]{Vasko15:jgr:cs}. The large $B_m$ makes from
almost all measured cross-tail $j_m$ the field-aligned current. The transverse current
magnitude usually several times smaller than the field-aligned current.
Interesting, due to different $B_m$ direction and different normal $\bm{n}$
direction, the measured field-aligned current can be positive or negative (from
event to event).

\begin{figure*}[h]
\centering
\begin{tabular}{c}
\includegraphics[width=0.7\textwidth, trim={5cm, 4cm, 7.cm, 3cm},
clip]{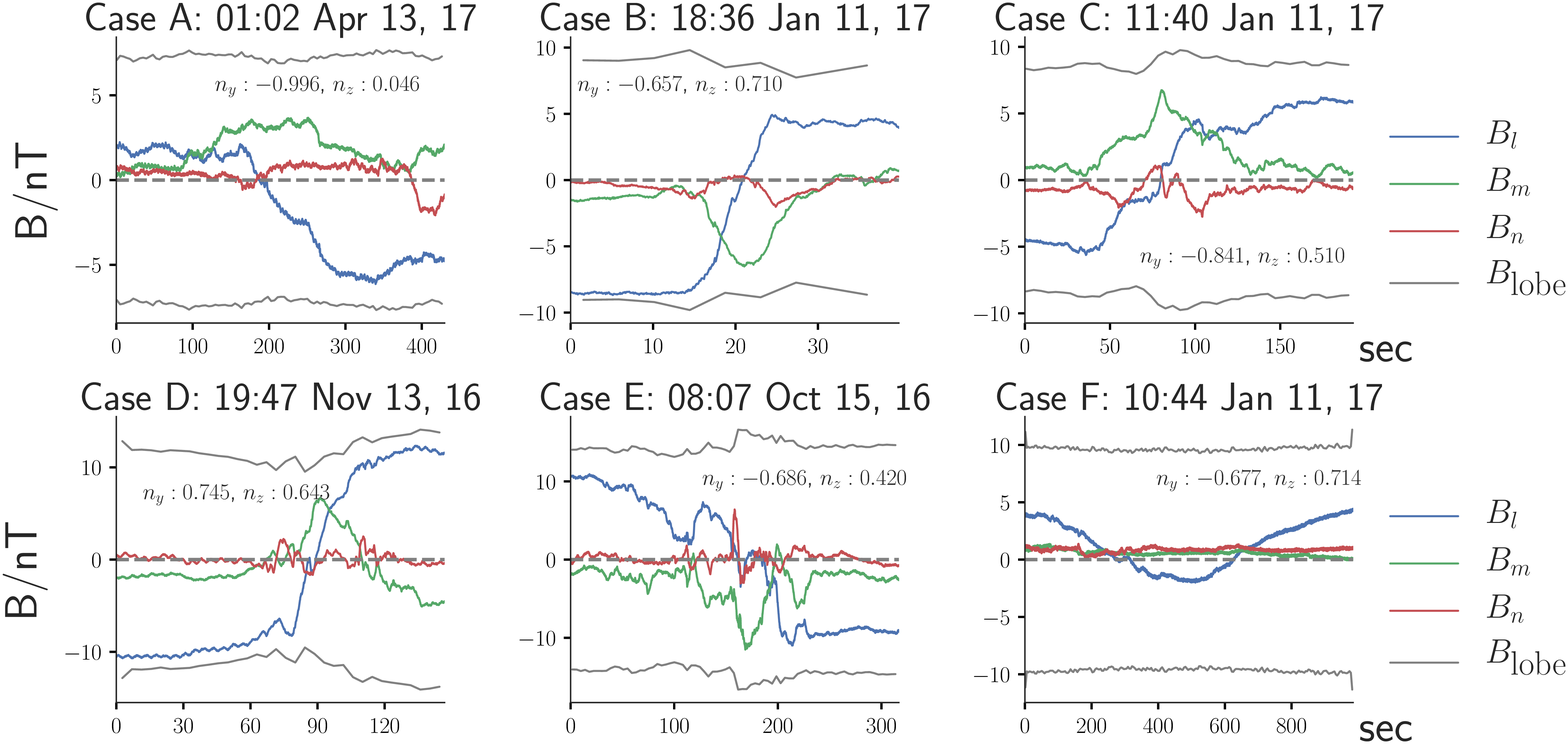} \\
\includegraphics[width=0.7\textwidth, trim={4.5cm, 3.5cm, 7cm, 3cm},
clip]{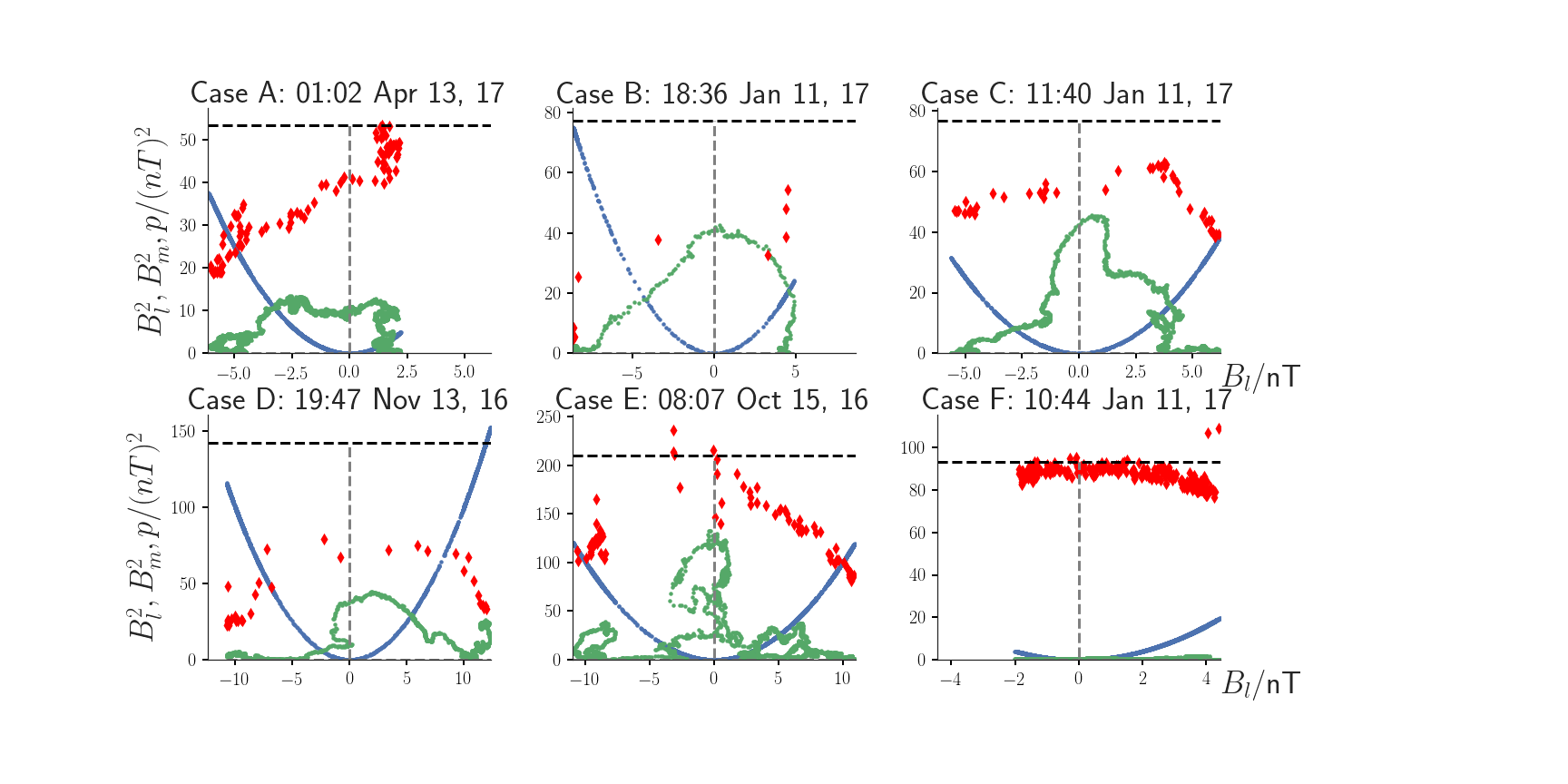}\\
\includegraphics[width=0.7\textwidth, trim={4.5cm, 3.0cm, 7cm, 3cm},
clip]{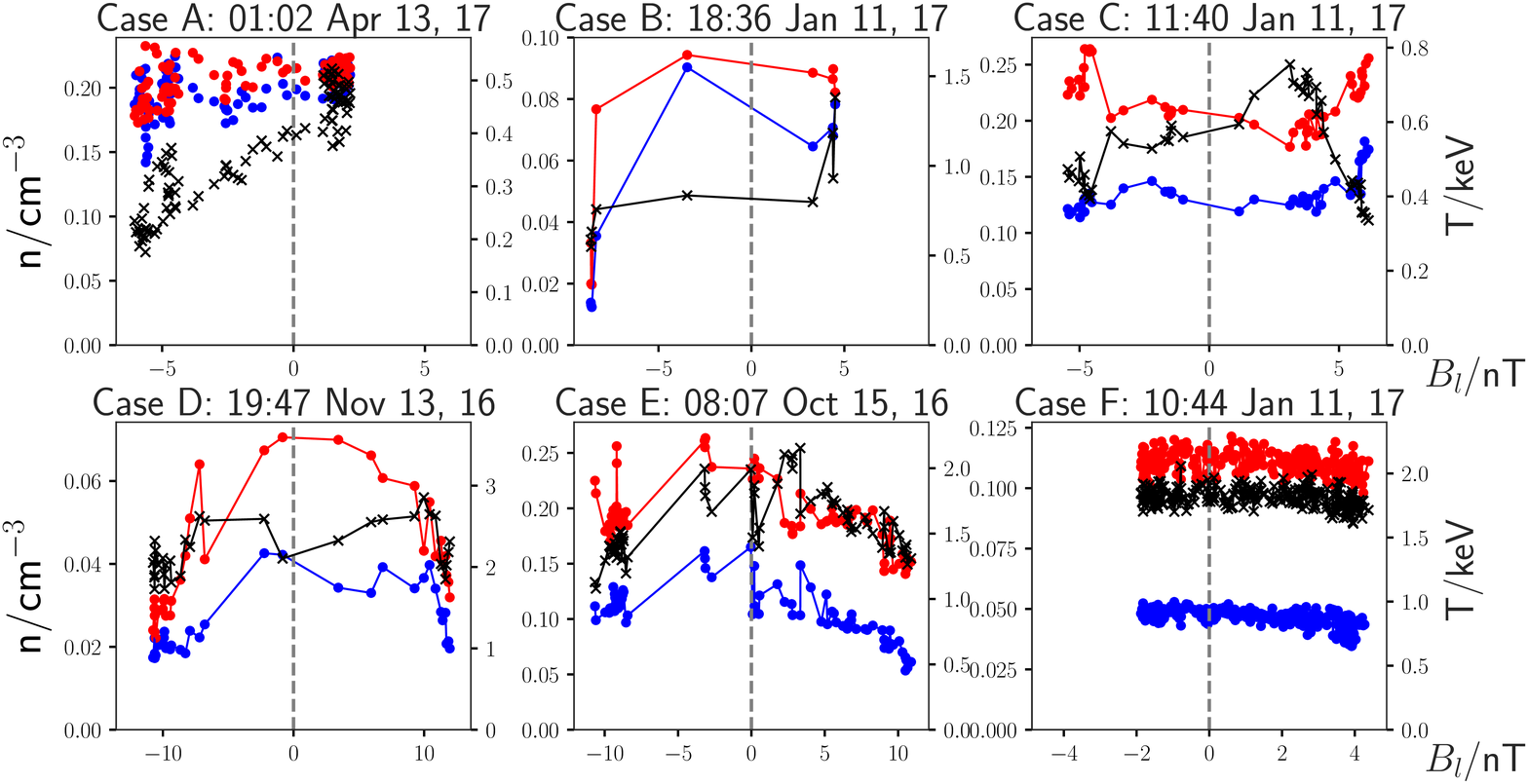} \\
\end{tabular}
\end{figure*}
\begin{figure*}[h]
\centering
\caption{Top six panels show time series of magnetic field components and lobe
magnetic field (grey curve). Middle six panels show pressure componnets
$B_l^2$(blue), $B_m^2$(green), and $p=2\mu_0k_Bn_e(T_i+T_e)$(red), the top
dashed lines indicate the average lobe pressure. Bottom six panels show
distributions of the density(black cross) and ion(red), electron(blue)
temperatures across the sheet. The electron temperature is multiplied by 5.}
\label{fig:B-t}
\end{figure*}

\section{Discussion and Conclusions}
We have demonstrated that a significant population of current sheets observed by
ARTEMIS at the lunar orbit consists of atypical current sheet configurations
with strong $B_m$ peak around the equatorial plane $B_l=0$. These current sheets
are tiled with the normal partially directed along dawn-dusk. The $B_m$ peak is
provided both by GSM $B_z$ and $B_y$ increase and contribute significantly to
the pressure balance. Existing of such specific current sheets can be related
to low plasma pressure (low density, temperature) which is insufficient to
balance the lobe (solar wind) pressure $\sim B_{\rm lobe}^2$. Similar current
sheet configurations were found in rarefied plasma of the Jupiters magnetotail
\citep[][]{Artemyev14:pss} and in cold plasma of Mars and Venus magnetotails
\citep{Rong15:venus, Artemyev17:jgr:mars}. In the Jupiter magnetotail current
sheets controlled by strong planet magnetic field, and observed field-aligned
currents represent the part of the global magnetosphere current system.
Instead,
Mars and Venus magnetotails form mainly by the solar wind magnetic field (due
to absence of regular planetary magnetic field) and currents sheets in these
magnetospheres resemble the solar wind rotational discontinuities
\citep[e.g.,][and references therein]{Paschmann13:angeo}. The Earth
magnetotail probed by ARTEMIS spacecraft represents some intermediate state
between Jupiter magnetotail filled by rarefied plasma but strongly connected to
planetary magnetic field and Mars/Venus magnetotail filled by cold plasma and
IMF field lines.
A closure of the intense field-aligned currents observed at lunar orbit is
unknown and needs to investigated with global models.  Independently
on the current closure, the observation of the statistically significant
population of that atypical current sheets at lunar orbit opens a
challenge for simulations and theories of the current sheet configuration.

Significant change in the current sheet configuration (in comparison with a simple diamagnetic configuration of the near-Earth tail) should result in change of the stability conditions and dynamical properties. Indeed, numerical simulations shows distinguishing features in magnetic reconnection initialization and development in current sheets with strong cross-tail field-aligned currents \citep[e.g.,][]{Zhou15:pop:reconnection, Wilson16:cs, Fan16:reconnection}. The primary region of the magnetotail reconnection ($\sim 30-40R_E$) is located right between the well investigated near-Earth current sheet with strong transverse cross-tail currents \citep[see statistics of Cluster results in][]{Runov06, Petrukovich15:ssr} and the lunar distant current sheet with strong cross-tail field-aligned currents. Therefore, current sheets at $\sim 30-40R_E$ can share properties of both these current sheet population and destabilization of this sheets can involve intensification of both transverse and field-aligned currents. This problem can be addressed by Magnetospheric Multiscale mission operating exactly within this region \citep{Burch16} and having chance to investigate the internal structure cross-tail field-aligned currents


\begin{acknowledgements}
We acknowledge NASA contract NAS502099 for the use of data from the THEMIS
Mission. We would like to thank the following people specifically C. W.
Carlson and J. P. McFadden for the use of ESA data, D.E. Larson and R.P. Lin for
the use of SST data, and K.H. Glassmeier, U. Auster, and W. Baumjohann for the
use of FGM data provided under the lead of the Technical University of
Braunschweig and with financial support through the German Ministry for Economy
and Technology and the German Aerospace Center (DLR) under contract 50 OC 0302.

The THEMIS data were downloaded from http://themis.ssl.berkeley.edu/.
\end{acknowledgements}

\bibliographystyle{agu}

\end{article}
\end{document}